\documentclass[12pt,aps,showpacs,superscriptaddress]{revtex4}
\usepackage{epsf}
\usepackage{graphicx}
\usepackage{dcolumn}
\usepackage{amsmath}

\begin{document}

\title{Nuclear effects on lepton polarization in charged-current
  quasielastic neutrino scattering}


\author{M. Valverde}
\affiliation{Departamento de Fisica Atomica, Molecular y Nuclear, 
          Universidad de Granada,
          Granada 18071, Spain}  
\author{ J.E. Amaro}
\affiliation{Departamento de Fisica Atomica, Molecular y Nuclear, 
          Universidad de Granada,
          Granada 18071, Spain}
\author{J. Nieves}
\affiliation{Departamento de Fisica Atomica, Molecular y Nuclear, 
          Universidad de Granada,
          Granada 18071, Spain}
\author{C. Maieron}
\affiliation{INFN, Sezione di Catania,
              Via Santa Sofia 64, 
              95123 Catania, Italy}

\begin{abstract}

We use a correlated local Fermi gas (LFG) model,
 which accounts also for long distance corrections of the RPA type
and final-state interactions, 
to compute the
polarization of the final lepton in charged-current quasielastic
neutrino scattering.  The present model has been successfully used in
recent studies of inclusive neutrino nucleus processes and muon
capture.  We investigate the relevance of nuclear effects in the
particular case of $\tau$ polarization in tau-neutrino induced
reactions for several kinematics of relevance for neutrino oscillation
experiments.

\end{abstract}




\maketitle


\section{Introduction}


The purpose of this paper is to investigate the importance of nuclear
 effects on the final lepton polarization in neutrino-induced
 charged-current (CC) inclusive reactions of the type $(\nu_l,l)$. In
 particular we present results for the tau polarization in
 $(\nu_\tau,\tau)$ reactions.  The interest of these studies on
 neutrino-nucleus reactions lies on their implications for the neutrino
 oscillations experiments \cite{fuk98a,fuk98b} (see \cite{Str06} and 
references therein for a recent review). Some of the
 experiments proposed to demonstrate the $\nu_\mu\rightarrow\nu_\tau$
 oscillation are expecting 
to detect the $\tau$ production signal
 through the $(\nu_\tau,\tau^-)$ or $(\overline{\nu}_\tau,\tau^+)$
 reactions \cite{icarus,Dra04}, 
among them the CNGS project \cite{cngs}, which will send a neutrino beam
from CERN to the Gran Sasso laboratory, 
where the ICARUS and OPERA detectors will
start taking data in the next few months. 
The $\tau$ decay particle distributions
 depend on the $\tau$ spin direction. Thus the theoretical information on the
 $\tau$ polarization will be valuable, since the expected number of
 $\nu_\tau$ events will not be large \cite{Kom02,Mig03,Ros05}. 
The study of $\tau$ polarization is also  needed, for instance, 
 in $\nu_\mu\rightarrow\nu_e$  oscillation experiments  
to disentangle $(\nu_e,e)$ events from background electron production 
following the $\nu_\mu\rightarrow \nu_\tau$ oscillation \cite{Hag03}.

The study of lepton polarization in $(\nu_l,l)$ and reactions is also
of theoretical interest since the polarization observables may display
peculiar sensitivities to the ingredients of the nuclear and reaction
models, different to the ones shown by the cross sections.  However
the optimal neutrino-energy regime to obtain partially polarized
leptons in non-longitudinal directions is limited. The reason is that
for high energy (compared to it mass) the final leptons are 100\%
polarized with negative helicity.  This is the case for the electrons
in $(\nu_e,e)$ reactions for most of the energies involved in
experiments.  In the case of muon production for some moderate
energies, a non negligible transverse polarization component, though
small, could be observed, and some examples will be shown below in this paper.
 More interesting is the aforementioned case
of tau leptons due to the large value of its mass. This lepton 
will be the main focus of this paper.

Previous studies of lepton polarization observables in neutrino
induced reactions focus mainly on $\tau$ polarization from nucleon
targets for a range of kinematics of interest
\cite{Hag03}--\cite{Kuz05}. 
Neutrino detectors are based on
neutrino-nucleus interactions (such as Argon in the ICARUS experiment),
and {\em a priori} one would need to evaluate the lepton
polarization observables in actual finite nuclei models. 
Genuine nuclear physics effects are usually neglected in these kind of
reactions, since relatively high energies are involved. However for low to
intermediate energy transfer nuclear effects do play a
role in the inclusive cross section of $(\nu,l)$ reactions 
and in general cannot be
neglected (Pauli blocking, finite size effects, long and short-range
correlations, final-state interactions, sub-nuclear degrees of freedom,
etc.).  This is well known from electron scattering studies, that usually
are the starting point of neutrino scattering models.  
For the same kinematics the $(e,e')$ reaction, and the $(\nu,l)$
reaction in the weak-vector sector are complementary to each other and
make use of the same dynamical ingredients. Only the weak-axial
current contribution makes a difference in the description of both
reactions, since it can excite different nuclear modes.

Only if a given model is able to reproduce to some extent the
available $(e,e')$ data, its predictions for neutrino reactions can be
considered reliable.  Different theoretical 
approaches along these lines on neutrino reactions in nuclei 
can be found  among the recent literature available
\cite{Hay00,Auer02,Mai03,Meu03,Ama05a,Ama05b,Cab05,Ama06,Mar06,Lei06}.

In this paper we describe the
$(\nu_l,l)$ reaction within the many-body framework of
refs. \cite{Nie04,Val06}. 
Our  model is an extension of previous studies on electron~\cite{GNO97},
photon~\cite{CO92}, and pion~\cite{OTW82,pion} interactions in nuclei. 
It is based on  
a Local Fermi Gas (LFG) description of the nucleus, 
accounting for Pauli blocking. 
The experimental
$Q$-values are used to 
enforce a correct energy balance of the reaction.
Additional nuclear effects, essential to describe $e$, $\gamma$ and $\pi$
 reactions, 
are built on top of the model, in particular: 
\begin{enumerate}
\item Coulomb distortion of the charged
leptons, 
\item Medium polarization, which
 is taken into account  
through the random-phase approximation (RPA),
\item  Final-state interaction (FSI). 
\end{enumerate}
The model was first applied to neutrino reactions in  Ref.~\cite{Nie04},
providing one of the best existing simultaneous
description of inclusive muon capture, $(\nu_\mu,\mu^-)$ and
$(\nu_e,e^-)$ reactions in $^{12}$C near threshold.  Inclusive muon
capture from other nuclei was also successfully described by the
model.  Apart from the description of the absorption of real photons
by nuclei~\cite{CO92}, the model describes rather well the
$(e,e^\prime)$ inclusive cross section of $^{12}$C, $^{40}$Ca and
$^{208}$Pb for different kinematics, not only in the QE region, but
also when extended to the $\Delta-$peak and the dip region
\cite{GNO97}. Recently the model has been extended to the neutral current 
sector, and one-nucleon knock-out reactions have been studied both for 
CC and neutral current driven processes \cite{Nie06}.
A special effort has also been paid in Ref.~\cite{Val06} to
reliably estimate the theoretical uncertainty of our model.

In this paper we extend the above model to the calculation of 
lepton polarization components in CC neutrino reactions.
To our knowledge, only one previous study exists analyzing nuclear
effects on lepton polarization \cite{Gra05b}.  This model basically
consists on a relativistic local Fermi gas (LFG) including nuclear
dynamical corrections such as some kind of relativistic RPA
correlations and an effective mass for the nucleons inside the
nucleus (this is an approximate way of taking into account the FSI). 
However the model of \cite{Gra05b}, first presented in
\cite{Gra03,Gra04}, has not been tested in other nuclear reactions
such as $(e,e')$. 

In Sect.\ 2 we introduce the formalism
and discuss the kinematics of the $(\nu_l,l)$ reaction for muon
and tau leptons in order to identify the most interesting cases. In
Sect.\ 3 we present results for differential cross sections and lepton
polarization components, and draw our conclusions.

\section{Formalism and Kinematics}

In this work we will study the inclusive neutrino-induced reaction
depicted in Fig.~1. A neutrino $\nu_l$ (or anti-neutrino $\overline{\nu_l}$)
with four momentum $K^\mu=(E_\nu,\vec{k})$ exchanges a $W$ boson with an atomic
nucleus with initial momentum $P^\mu=(M_i,\vec{0})$, and a polarized 
lepton $l^-$ (or
$l^+$) is detected with four-momentum $K'{}^\mu=(E'_l,\vec{k'})$. In the
inclusive reaction the final hadronic state is not detected. In this work we
deal with the quasielastic channel, where the main contribution is due to
one-nucleon emission.

\begin{figure}[th]
\begin{center}
\includegraphics[scale=0.8, bb= 150 460 520 650]{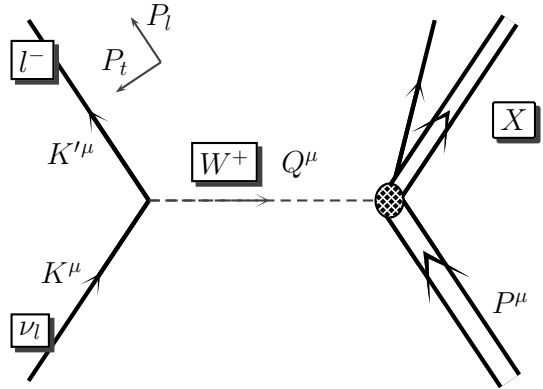}
\caption{
Kinematics of the reaction, where we show the 
scattering plane and the 
directions of the two lepton polarization components $P_l$ and $P_t$.
\label{kin} }
\end{center}
\end{figure}

We write the unpolarized differential cross section as \cite{Nie04}
\begin{equation} \label{sigma}
\Sigma_0 \equiv \frac{d^2\sigma_{\nu l}}{d\Omega'dE^\prime_l} =
\frac{|\vec{k}^\prime|G^2M_i}{2\pi^2} F,
\end{equation}
where $G$ is the Fermi weak coupling constant, 
$\Omega'$ is the solid angle of the final lepton, and the quantity 
 $F$ has been defined as 
\begin{eqnarray}
F & = & 
\left( 2W_1 +\frac{m_l^2}{M_i^2}W_4\right)
(E_l^\prime-|\vec{k}^\prime|\cos\theta)
+ W_2(E_l^\prime+|\vec{k}^\prime|\cos \theta) 
\nonumber\\
&&\mbox{}
- W_5\frac{m_l^2}{M_i} \mp \frac{W_3}{M_i}
\left(  E_{\nu}E_l^\prime+|\vec{k}^\prime|^2
       -(E_{\nu}+E_l^\prime)|\vec{k}^\prime|\cos\theta
\right)
\end{eqnarray}
resulting from 
 the usual contraction between the leptonic and hadronic tensors. 
Here $\theta$ is the angle between $\vec{k}$ and $\vec{k'}$, and 
the $\mp$ sign in the last term 
correspond to the case of neutrino or anti-neutrino scattering. 
Finally the structure functions $W_i$  are defined in the hadronic tensor as
\begin{equation}
\frac{W^{\mu\nu}}{2M_i} = - g^{\mu\nu}W_1 + \frac{P^\mu
  P^\nu}{M_i^2} W_2 + {\rm i}
  \frac{\epsilon^{\mu\nu\gamma\delta}P_\gamma q_\delta}{2M_i^2}W_3 +  
\frac{q^\mu  q^\nu}{M_i^2} W_4 + \frac{P^\mu q^\nu + P^\nu q^\mu}
{2M_i^2} W_5
\end{equation}
with $\epsilon_{0123}= +1$ and the metric $g^{\mu\nu}=(+,-,-,-)$.

\begin{figure}[th]
\begin{center}
\includegraphics[scale=0.8,  bb= 50 450 540 790]{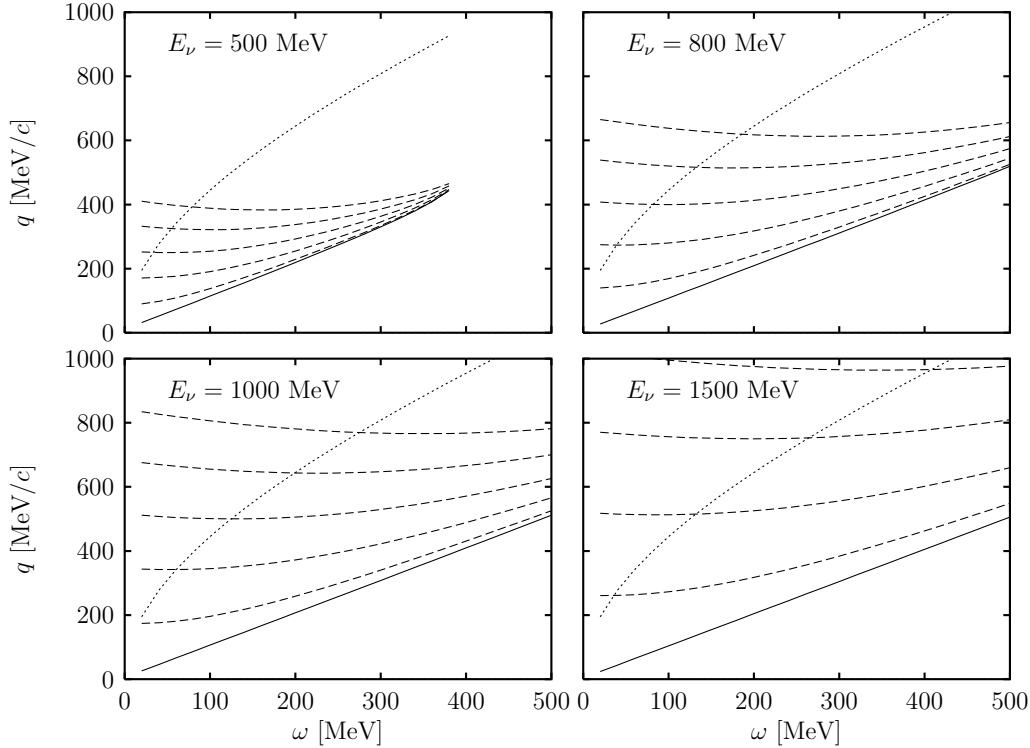}
\caption{Allowed kinematics for $(\nu_\mu,\mu)$ reaction in the
$(\omega,q)$ plane. The dotted line is the center of the quasielastic
peak $\omega=(q^2-\omega^2)/2m_N$. The other lines show in each panel the value
of $q$ as a function of $\omega$ for a fixed value of the neutrino
energy, and for several values of the scattering angle $\theta$.
Starting with the solid line for $\theta=0$, the angles from down to
up correspond to $\theta=10^{\rm o}, 20^{\rm o}, 30^{\rm o}, 40^{\rm
o}$ and $50^{\rm o}$.
\label{kmuon1} }
\end{center}
\end{figure}

We assume that the final lepton polarization is measured in the
direction defined by the vector $s^\mu$ verifying $s^2=-1$. The
polarized differential cross section (\ref{sigma}) can be written as
\begin{equation}
\Sigma = \frac12\Sigma_0\left(1+s_\mu P^\mu\right)
\end{equation}
where $\Sigma_0$, given in Eq.~(1),
 is the cross section corresponding to unpolarized leptons and
$P^\mu$ is the polarization vector.  The relevant components of the
polarization vector in the laboratory system are defined in Fig. 1 and denoted
$P_l$ (longitudinal, in the direction of $\vec{k'}$), and $P_t$ (transverse to
$\vec{k'}$ and contained in the scattering plane).  
Working within the standard model, as we do in the present work, 
it can be shown that the
polarization component perpendicular to the scattering plane (Fig. 1) is zero.
\cite{Kuz04,Llew71}. 
The expressions for the two polarization components, $P_l$ and $P_t$, in terms
of the hadronic structure functions $W_i$ are as follows
\begin{eqnarray}
P_l &=& 
\mp \left\{ 
\left(2W_1 - \frac{m_l^2}{M_i^2}W_4\right)
  (|\vec{k}^\prime|-E_l^\prime\cos\theta)
    + W_2(|\vec{k}^\prime|+E_l^\prime\cos\theta)     
    - W_5\frac{m_l^2}{M_i}\cos\theta  
\right. \\ \nonumber
& & \left. 
    \mp \frac{W_3}{M_i}((E_\nu+E_l^\prime)|\vec{k}^\prime|
           -(E_\nu E_l^\prime+|\vec{k}^\prime|^2)\cos\theta) 
\right\} /F \\
P_t &=&
 \mp m_l\sin\theta
   \left(2W_1 - W_2 - \frac{m_l^2}{M_i^2}W_4 + W_5\frac{E_l^\prime}{M_i}
    \mp W_3\frac{E_\nu}{M_i}\right)/F
\end{eqnarray}
Note that the transverse polarization $P_t$ is proportional to the lepton
mass. The larger values of this polarization component are then expected for
tau leptons, while it will be negligible for electrons at the intermediate
energies of interest for neutrino reactions.

\begin{figure}[th]
\begin{center}
\includegraphics[scale=0.8,  bb= 50 450 540 790]{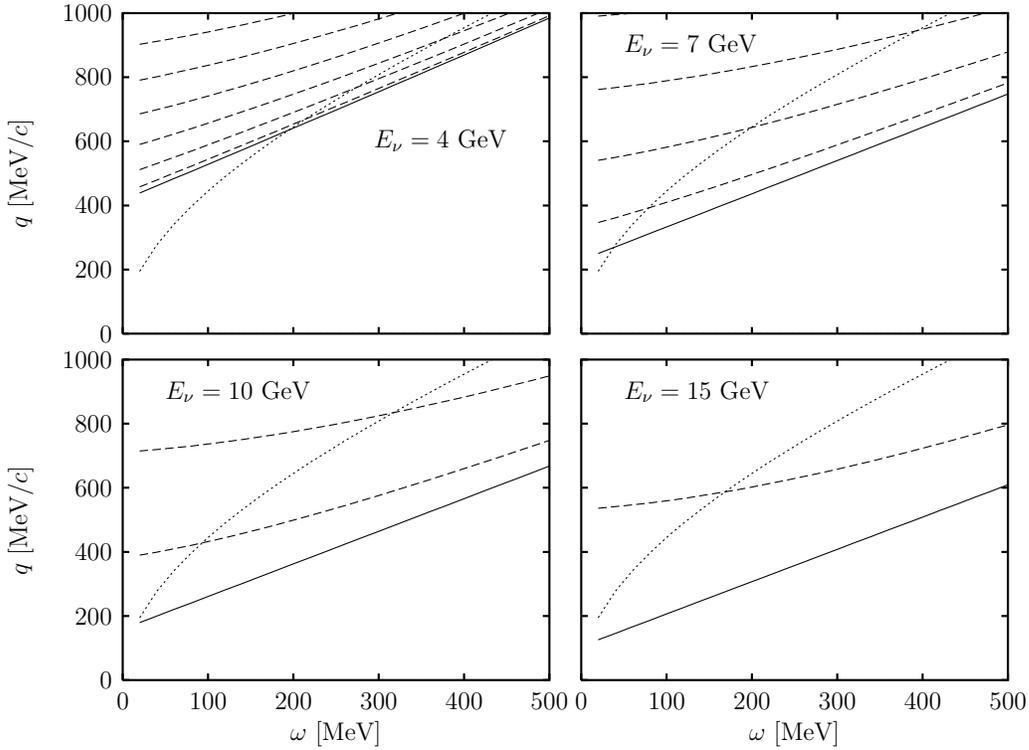}
\caption{Allowed kinematics for $(\nu_\tau,\tau)$ reaction 
  in the $(\omega,q)$ plane. The dotted line is the center of the quasielastic
  peak $\omega=(q^2-\omega^2)/2m_N$. 
The other lines show the value of $q$ as a function
  of $\omega$ for several fixed values of the scattering angle $\theta$.
  Starting with the solid line for $\theta=0$, the angles from down to up
  correspond to $\theta=2^{\rm o}, 4^{\rm o}, 6^{\rm o}, 8^{\rm o}, 10^{\rm
    o}$ and $12^{\rm o}$.
\label{ktau1} }
\end{center}
\end{figure}

Let us begin with a  discussion on the kinematics of the reaction in order to
identify the cases of interest. For a fixed value of the neutrino 
energy, $E_\nu$, and from the definition of the four-momentum
transfer, $Q^\mu=K^\mu-K'{}^\mu=(\omega,\vec{q})$, one easily obtains
the modulus of the three-momentum transfer $q\equiv|\vec{q}|$,
\begin{equation}
q^2=\omega^2-m_l^2+2E_\nu(E_\nu-\omega)-2E_\nu\sqrt{(E_\nu-\omega)^2-m_l^2}
\cos\theta.
\end{equation}
For a fixed value of the scattering angle, the above equation gives $q$ as a
function of $\omega$, while the case $\theta=0$ leads to a parametrized
curve corresponding to the the boundary of the allowed kinematical region in
the $(\omega,q)$ plane. The identification of this region will help in the
present work because we deal with a model describing the reaction around the
quasielastic (QE) peak defined by $\omega=(q^2-\omega^2)/2m_N$, with $m_N$ the
nucleon mass. Therefore we must choose kinematics for which the QE peak lies
inside the allowed region. Some examples are shown in Figs. 2 and 3 for muon
and tau leptons, respectively.

In Fig.~2 the muon case is displayed for four values of the neutrino energy
ranging from $E_\nu=500$ to 1500 MeV. In each panel we show the possible
kinematics in the $(\omega,q)$-plane for several values of the scattering
angle, $\theta=10^{\rm o}, 20^{\rm o}, 30^{\rm o}, 40^{\rm o}$ and $50^{\rm
  o}$. The lower solid lines correspond to the limit value $\theta=0$,
setting the lower boundary of the allowed kinematical region.  The maximum
value of the energy transfer $\omega_{\rm max}=E_\nu-m_l$ sets the right end
of the boundary.  In the same plots we also show the position of the maximum
of the QE peak $\omega=(q^2-\omega^2)/2m_N$ with dotted lines.  Note that in
all cases the QE peak region is inside the allowed region for $(\nu_\mu,\mu)$
reactions. In the plots we only show the low to intermediate energy transfer
part $\omega<500$ MeV since we are interested in the non relativistic regime
where our nuclear model is safely applicable. This also means that the
momentum transfer $q$ must be below 600 or 700 MeV/c, since relativistic
corrections at these values start to be important. Although our
nuclear model is based on a fully relativistic description of the Fermi gas,
the RPA and FSI corrections are non relativistic. Therefore for this work we
are forced to kinematics with low to intermediate energy and momentum
transfer. The plots in Figs 2 and 3 are very useful to this end since they
give us a clear picture of the kinematical changes when the scattering angle
is increased.

The case of $\tau$ leptons is considered in Fig. 3 for neutrino energies going
from $E_\nu=4$ to 15 GeV.  This time the scattering angles shown are
$\theta=0, 2^{\rm o}, 4^{\rm o}, 6^{\rm o}, 8^{\rm o}, 10^{\rm o}$ and
$12^{\rm o}$.
In contrast to the muon case, for tau leptons the QE region is forbidden
for some kinematics. In particular for low neutrino energy and low scattering
angles. For instance, for $E_\nu$=4 GeV the maximum of QE peak is below the 
$\theta=0$ curve for $\omega$ smaller than 200 MeV. In order to cover this
region, one should go to momentum transfers well above 800 MeV/c, where
relativity would play a role. Therefore we must go to larger 
neutrino energies, above $\sim 7$ GeV, in order to cover the QE region 
for small values of $q$ and $\omega$.
Moreover, for large values of $E_\nu$ the momentum transfer strongly increases 
with the scattering angle, and then we force it to fall in the 
few degrees region to guarantee non relativistic kinematics.

\section{Results and concluding remarks.}

\begin{figure}[th]
\begin{center}
\includegraphics[scale=0.8,  bb= 50 400 540 800]{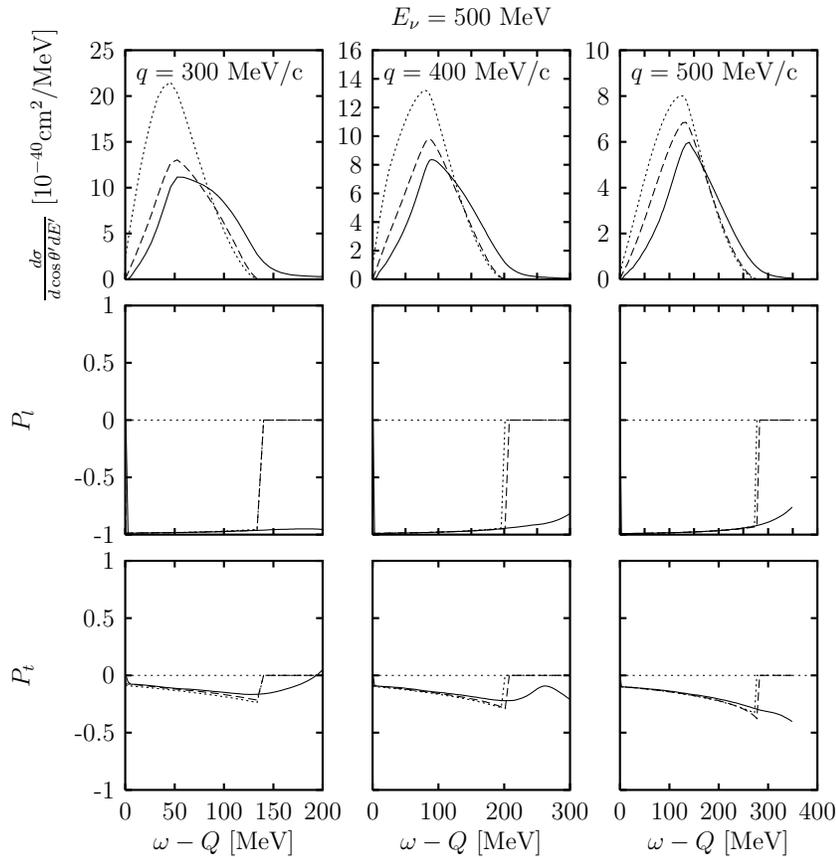}
\caption{
  Differential cross section and polarization components for the
  $^{40}$Ar$(\nu_\mu,\mu)$ reaction as a function of the energy transfer minus
  the experimental $Q$-value, and for three values of the momentum transfer
  $q=300, 400$, and 500 MeV/c. The incident neutrino energy is fixed to 500
  MeV.
\label{muon1} }
\end{center}
\end{figure}

Our results showing the effects of the different corrections
implemented into a nuclear model are summarized in Figs. 4--8.  With
dotted lines we show the results of our model without RPA and without
FSI corrections.  Hence these results account basically for the
impulse approximation with neutrino-nucleon interaction of $V-A$
type. We use the Galster parameterization for the nucleon vector form
factors and a dipole dependence for the axial form factor, while the
pseudo-scalar form factor is related to the later by the partially
conserved axial current (PCAC) hypothesis.  These results include also
the following effects:
\begin{enumerate}
\item  Pauli blocking, through a LFG description of the nucleus. This implies
  an additional dependence on the experimental nuclear density. 
\item Correct energy balance of the reaction using   experimental $Q-$values. 
In the figures we show $q$ as a function of
  $\omega$ minus the $Q$ value.
\item   Coulomb distortion of the charged leptons. 
\end{enumerate}
More details on the parameters of CC current and on the model 
are given in our previous works \cite{Nie04,Val06}.

\begin{figure}[tbh]
\begin{center}
\includegraphics[scale=0.8,  bb= 50 400 540 800]{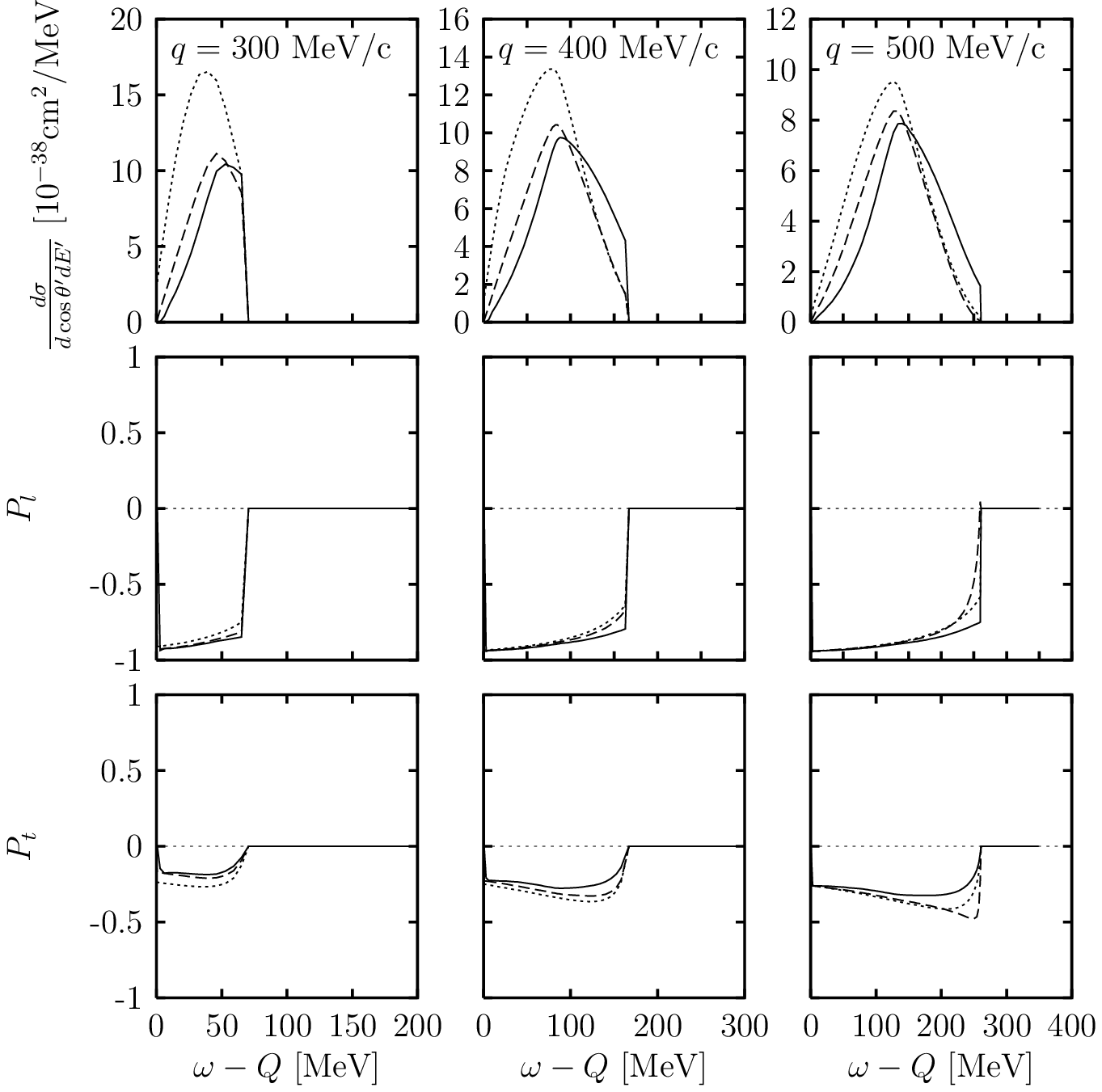}
\caption{
The same as Fig. \ref{muon1} for tau neutrinos of 7 GeV.
\label{tau1} }
\end{center}
\end{figure}

\begin{figure}[th]
\begin{center}
\includegraphics[scale=0.8,  bb= 50 400 540 800]{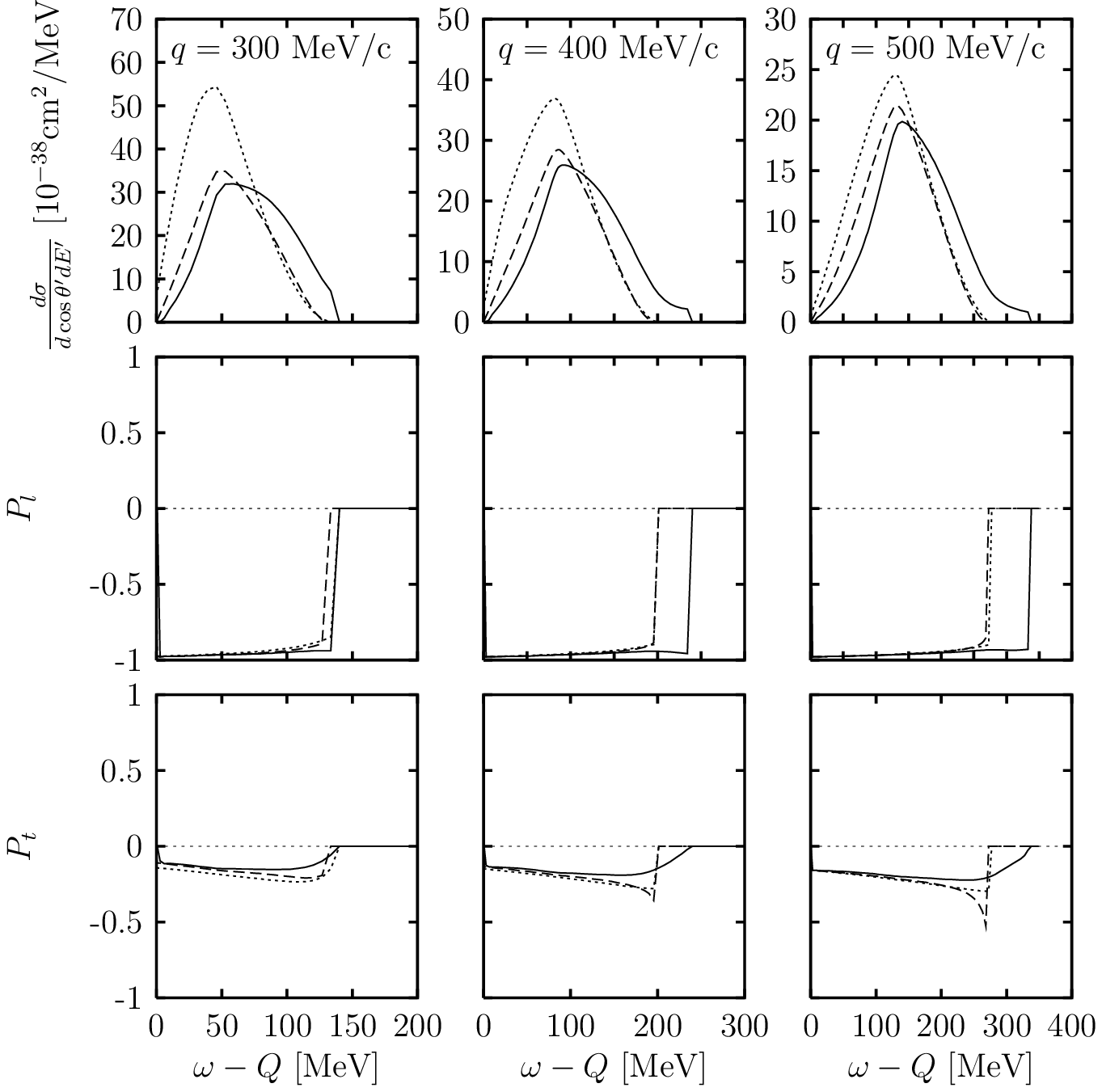}
\caption{
The same as Fig. \ref{tau1} for tau neutrinos of 10 GeV.
\label{tau2} }
\end{center}
\end{figure}

In the same plots we show with dashed lines the results including RPA
corrections, that take into account modification of the nuclear medium
through interactions between particle-hole and $\Delta-$hole
excitations.  We use an effective nucleon--nucleon interaction, with
pion and rho exchange in the vector-isovector channel, and corrections
due to short-range correlations.  We refer the reader to
Ref. \cite{Nie04} for details about the RPA series and for explicit
expressions of the nuclear structure functions $W_i$ entering in the
CC neutrino cross section.

Finally we show with solid lines the results with the full model, including
also the FSI.  We account for relevant reaction mechanisms where two nucleons
participate, by dressing the nucleon propagators in the nuclear medium. In
particular these effects change the dispersion relation of the nucleon, which in
some works is taken into account by the over-simplified method of introducing
an effective mass for the nucleon \cite{Gra05b}.

We start the discussion with Fig. 4 where we show typical results for
the $(\nu_\mu,\mu)$ reaction.  In the figure we show the relevant
observables, namely the differential cross section and the two
polarization components, $P_l$ and $P_t$ in the laboratory system.
The neutrino energy is $500$ MeV, and three values of the momentum
transfer $q=300$, 400 and 500 MeV/c are considered.  These values of
the momentum transfer correspond to lepton scattering angles of
30$^{\rm o}$ and above. The cross section shows the typical behavior
of the QE peak and we can see how the different nuclear effects modify
this observable. First the RPA produces a big reduction and small
shift of the peak (compare the dotted to the dashed lines). The
quenching of the cross section is large for low momentum transfer
(almost a 50\% reduction at the peak for $q=300$ MeV/c) and diminishes
with $q$ (below 15\% for $q=500$ MeV/c). On the other hand the FSI
produces a further reduction of the RPA results and an important
enhancement for high energy transfer.  This a consequence of the
modification of the nucleon dispersion relation in the medium
through the dressed nucleon propagator inside the nucleus.

Concerning the polarization observables, we see that the longitudinal
component $P_l$ is very close to $-1$ for the three kinematics, while
there is a small but appreciable transverse component (around $-0.2$),
quite independent of $q$. The inclusion of the RPA does not change
these results.  The reason is that the polarization components are
obtained as a ratio between linear combinations of nuclear
structure functions and the RPA changes similarly numerator and
denominator. The same can be said for the FSI effects, except for
the high-$\omega$ region. The LFG is unable to
contribute to the high energy tail of the cross section, 
while the model with FSI is able to describe this region where, however, the 
QE cross section is rather small.

We have generated results for higher values of the muon-neutrino
energy corresponding to the kinematics of Fig. 2. However in all the
remaining cases the transverse polarization component computed is
negligible and the muon can be considered polarized with negative
helicity.

An example of the results found for tau leptons is displayed in
Fig. 5, where we show the $(\nu_\tau,\tau)$ cross section and
polarization observables for the same values of $q$ as in Fig. 4, 
but this time for
$E_\nu=7$ GeV. By inspection of Fig. 3 we see that for the three values of
$q$ considered the maximum of the QE lies inside the allowed
kinematical region. The case $q=300$ MeV/c is closer to
the boundary, since the cross section  suddenly ends soon above the
maximum, while for higher values of the momentum
transfer the allowed region extends to higher energy transfer. The
effects seen here over the cross section due to RPA and FSI are similar
to the muon case studied above, with the exception of the missing high
energy tail that lies now in the forbidden region.

\begin{figure}[tb]
\begin{center}
\includegraphics[scale=0.8,  bb= 50 400 540 800]{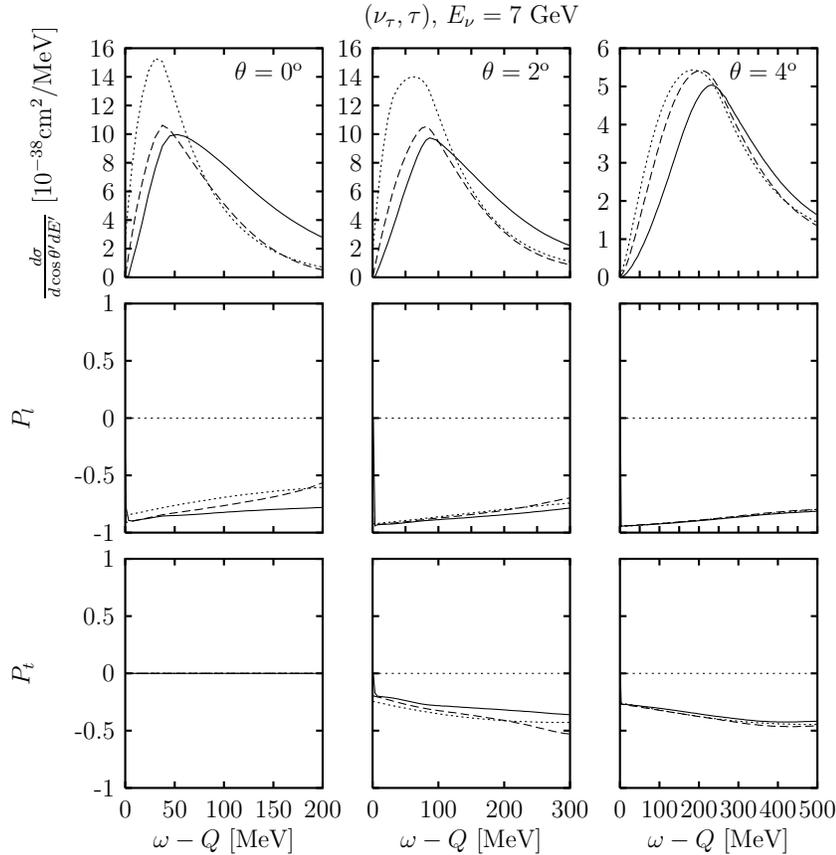}
\caption{ The same as Fig. \ref{tau1} fixing the scattering angle
$\theta$ instead of the momentum transfer $q$. Three values of
$\theta=0^{\rm o}, 2^{\rm o}$ and $4^{\rm o}$ are shown.
\label{tau4} }
\end{center}
\end{figure}

 Concerning the polarization components, the longitudinal one is still
negative but now is well above $-1$. On the other hand, an appreciable
$P_t$ component appears, taking values ranging from $\sim -0.2$ to $\sim
-0.25$. Thus the magnitude of $P_t$ increases with $q$, and, 
as in the muon case, the RPA and FSI effects are rather small 
on the polarization observables. 

A more clear scene with similar results can be seen in Fig. 6 for
$E_\nu=10$ GeV, corresponding to the kinematics of the third panel in
Fig. 3.  In this case the QE peak is well inside the allowed region
for the three values of $q$. The polarization component $P_l$ is now
closer to $-1$, while the magnitude of $P_t$ is smaller than in the
former case.  Results for $E_\nu=15$ GeV (not shown) indicate that
$P_l$ almost reaches  the limit value $\sim -1$, while $P_l$ becomes very
small for high neutrino energy, as expected.

Another kind of plot of interest for ongoing neutrino experiments is
shown in Fig.\ 7 for $\tau$ neutrinos with energy of 7 GeV.  This time
we fix the scattering angle instead of the momentum transfer.
Therefore by changing $\omega$ we are always inside the allowed
kinematical region, running along some of the curves shown in the
second panel of Fig. 3, never crossing the boundary, and the cross
section is defined for every value of $\omega$ shown in the plots.
Results are shown for small angles, $\theta=0, 2^{\rm o}$, and $4^{\rm
o}$, in order to reach not too high values of the momentum transfer.
For these angles the maximum of the QE peak is crossed at $q\sim 300,
400$, and 600, respectively. Except for $\theta=4^{\rm o}$ at the high
$\omega$ tail where the $q$-values are perhaps too high, and a word of
caution is needed since some important relativistic corrections are
expected, we can safely trust the results for lower $\omega$-values
where $q$ is moderately low.  Again we can see the important reduction
of the cross section due to RPA correlations for low $\omega$. The
magnitude of this reduction decreases with the scattering angle. The
FSI produces a reduction for low $\omega$ and an increase for high
values of the energy transfer. The polarization component $P_l$ is
more or less independent on the angle and takes values between -0.9
and -0.8.  The transverse polarization $P_t$ is zero for $\theta=0$ by
definition, Eq. (6), while it is not negligible at all for the
remaining angles, taking values between $\sim -0.2$ and -0.5.
Concerning the RPA and FSI corrections on these polarization
observables, the effect is found to be  negligible for 
low energy.  However
at the high energy tail and for $\theta=0$ the net RPA+FSI effect
makes  $P_l$ to change from $\sim -0.6$ to $\sim -0.8$.
A less important, but still appreciable change is found in $P_t$ 
for $\theta=2^{\rm o}$. For higher values of the scattering angle 
the effect is again negligible.

\begin{figure}[tb]
\begin{center}
\includegraphics[scale=0.8,  bb= 50 520 540 780]{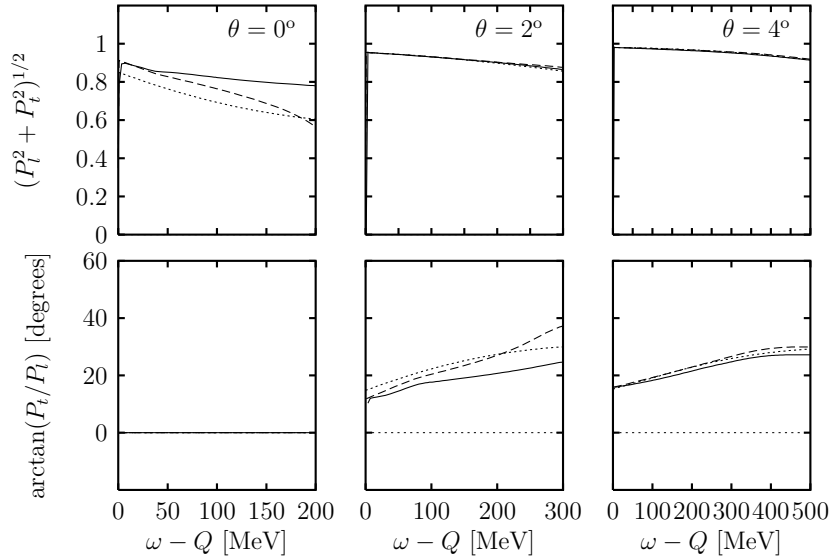}
\caption{ Modulus and angle of the polarization vector with respect to the 
$-l$ direction for tau neutrinos of energy 7 GeV, and for three scattering 
angles, corresponding to the kinematics of fig. \ref{tau4}.
\label{tau7} }
\end{center}
\end{figure}

To end the discussion, in Fig. 8 we show another representation of
polarization observables, namely, the total polarization, defined as the 
modulus of the polarization
vector 
\[ |\vec{P}|=\sqrt{P_l^2+P_t^2},
\]
 and the angle  with respect to minus the longitudinal
direction ($-\vec{k'}$),
\[
\Theta\equiv\arctan(P_t/P_l).
\]
The total polarization $|\vec{P}|$ takes values between 0.8 and 0.9
for $\theta=0$ and increases with the scattering angle. For $\theta=0$
the angle $\Theta$ is also zero, 
meaning that the polarization vector points to the
$-\vec{k'}$ direction, without transverse components. In this case the
total polarization has the meaning of fraction of particles with
negative helicity.  A value different from unity means that the
interaction with the nuclear target produces a small fraction of tau
leptons with {\em positive helicity}.  A value close to one means that
a large percentage of the leptons exit with the spin pointing to the
same direction. The most probable direction of the lepton spin is
determined by the angle $\Theta$. For $\theta=2^{\rm o}$ this angle is
between $\sim 15$ and $25^{\rm o}$, and slightly increase for
$\theta=4^{\rm o}$.  The results of Fig. 8 are showing a non
negligible increase of the total polarization due to RPA+FSI effects,
for $\theta=0$. These effects are negligible over this observable for
$\theta=2$ and $4^{\rm o}$. On the other hand, for $\theta=2^{\rm o}$
we find an appreciable reduction of the polarization angle $\Theta$
due to RPA+FSI, that again is negligible for higher scattering angles.

Summarizing this work, we have computed the cross section and
polarization observables for neutrino induced CC reactions in nuclei
at the QE peak. We have focused this work on the case of $\tau$
leptons of interest for neutrino oscillation experiments.  The RPA
plus FSI nuclear effects have been evaluated for intermediate energy
and momentum transfer. These corrections are essential in the cross
section, especially for low energy, but are partially reduced, due to
cancellations, on the polarization observables.  However, we have identified 
 some particular kinematics, at very low scattering angles, where 
these nuclear corrections are of some importance to 
determine the correct magnitude and angle of the polarization vector.

\begin{acknowledgments}

This work was supported by DGI and FEDER funds, contract
FIS2005-00810, by the EU Integrated Infrastructure Initiative Hadron
Physics Project contract RII3-CT-2004-506078 and by the Junta de
Andaluc\'\i a.

\end{acknowledgments}

\end{document}